\documentclass[a4paper,superscriptaddress,showpacs,nofootinbib,preprintnumbers,amsmath,amssymb]{revtex4}

\usepackage[dvipdfm]{graphicx}
\usepackage{amssymb}
\usepackage{amsmath}
\usepackage[nohead]{geometry}
\usepackage{parskip}
\usepackage{wick}
\usepackage{epsfig}

\begin{document}

\bibliographystyle{prsty}

\title{Gravity Induced Chiral Condensate Formation and \\ the Cosmological Constant}
\author{Stephon Alexander}
\author{Deepak Vaid}
\affiliation{Department of Physics,\\
Institute for Gravitational Physics and Geometry,\\
The Pennsylvania State University,\\
104 Davey Lab, University Park, PA,16802, U.S.A \\
IGPG-06/8-4 }

\date{\today}

\begin{abstract}
It is well known that the covariant coupling of fermionic matter to gravity induces a
four-fermion interaction. The presence of this term in a homogenous and isotropic space-time
results in a BCS-like Hamiltonian and the formation of a chiral condensate with a mass gap. We
calculate the gap ($\Delta$) via a mean-field approximation for minimally coupled fermionic fields
in a FRW background and find that it depends on the scale factor. The calculation also yields a
correction to the bare cosmological constant ($\Lambda_0$), and a non-zero vev for
$<\psi^\dag\psi>$ which then behaves as a scalar field. Hence we conjecture that the presence of
fermionic matter in gravity provides a natural mechanism for relaxation of the $\Lambda_0$ and
explains the existence of a scalar field from (almost) first principles.
\end{abstract}


\maketitle

\section{Introduction}

Ever since the BCS theory of superconductivity has been discovered, the phenomenon of Cooper
pairing has played a seminal role across a wide range of physics, including Pion formation,
Technicolor and QCD at high densities.  A Cooper pair requires some necessary conditions:
\begin{itemize}
\item A Fermi surface. \item Screening resulting in an attractive interaction between fermions.
\item A relevant four-fermion interaction.
\end{itemize}

Another important aspect of the BCS theory is that it signifies that the perturbative vacuum with
respect to perturbative phonon or vector boson exchange is unstable; very weak attractive
interaction drives the system to a lower energy non-perturbative ground state.  In the context of
general relativity graviton exchange between fermions is a ripe setting to ask whether or not a BCS
condensate can form. This possibility may have consequences, especially for the inflationary
paradigm and the cosmological constant problem since the idea that the vacuum is unstable with
respect to graviton exchange between fermions can pave a way to solving the cosmological constant
problem.  In this paper we demonstrate for the first time that gravity naturally incorporates a BCS
condensate in a general, $\Lambda$ dominated, FRW space-time.  In the context of inflation this
condensate can play the role of the inflaton field.  We also show that the gap introduces a
non-perturbative cancelling correction to the cosmological constant which is consistent with the
expectations of the authors as a possible path towards resolving the cosmological constant problem.
Other approaches to the dark energy problem as a condensate has been proposed in the past but a concrete microphysical mechanism
has been lacking \cite{robert,nima}.  We hope that this work may be useful in providing the correct microphysics underlying the dark energy
problem in terms of condensates.

Recently \cite{Rovelli_Perez_1} is was shown that gravity in the presence of a Dirac field induces
a non-zero torsion. This torsion turns out to be proportional to the axial current, $J_{\mu5}$.
Inserting the expression for the torsion back into the first-order action we find a new interaction
term which is proportional to the square of the axial current and also has a dependance on the
Immirzi parameter.\footnote{This four-fermion interaction is not new. As far back as 1922 Cartan
proposed that a correct theory of gravity should also contain torsion.} Such a four-fermi
interaction is well-known to cause the formation of a chiral condensate. As a consequence
$<\psi^\dag\psi>$ develops a non-zero vev and the resulting theory has a mass gap $\Delta$. We also
find a negative contribution to the cosmological constant $\Lambda_0$ from the fermionic
condensate.

The paper is arranged as follows. In Section 2 we show how the presence of a Dirac term in the
first-order action for fermions coupled to gravity, induces the four-fermion interaction. In
Section 3 we do the (3+1) decomposition of the resulting Lagrangian and find the Hamiltonian by
performing a Legendre transform. This allows us to identify the diffeomorphism, hamiltonian and
gauge constraints of the theory. It is the hamiltonian constraint which is responsible for dynamics
and we concentrate on it. In Section 4 we write down the hamiltonian constraint for a FRW metric.
We then quantize the fermion field, while leaving the background metric classical. In Section 5 we
exhibit the Boguliubov transformation on the fermionic ladder operators which is a necessary step
in the BCS calculation\footnote{The gap can also be determined via a variational method, however
the Boguliubov transformation is simpler and more instructive}. In Section 6 we diagonalize the
matter hamiltonian by applying the Boguliubov transformation and then find the gap equation. We
find that the gap has a dependence on the scale factor and acts to negate the cosmological constant
term in the hamiltonian constraint. Finally we discuss our results and mention avenues for future
research.

\section{Torsion and the four-fermi interaction}
Our starting point is with the Holst action for General Relativity with a cosmological constant,
coupled to fermions.  We will calculate the four-fermion interaction induced by Torsion and write
the action in Hamiltonian form.  The action will be symmetry reduced and after all of the
constraints are identified we will show that the fermionic Hamiltonian is a many-body BCS
Hamiltonian.  Finally we will diagonalize the Hamiltonian and calculate the energy gap.

First, it is convenient to introduce our conventions.  Lowercase greek letters $\mu,\nu,...$ stand
for four dimensional spacetime indices $1..4$. Lowercase latin letters denote spatial indices on
$\Sigma$. Uppercase latin $I,J,...$ denote internal indices $1..4$. Lowercase latin letters denote
internal indices $1..3$.

The action for gravity coupled with massless fermions is:

\begin{equation}\label{action}
    S[A,e,\Psi] = S_{H} + S_{D}
\end{equation}

where $S_{H}$ is the Holst action and is equivalent to the metric formulation of general
relativity:

\begin{equation}\label{holst_action}
    S_{H} = \frac{1}{2\kappa}\int d^{4}x\,e\,e^{\mu}_{I}e^{\nu}_{J}F^{IJ}_{\mu\nu} -
    \frac{1}{2\kappa\gamma}\int d^{4}x\,e\,e^{\mu}_{I}e^{\nu}_{J}\star F^{IJ}_{\mu\nu}
\end{equation}

and $S_{D}$ is the action for fermions:

\begin{equation}\label{dirac_action}
    S_{D} = \frac{i}{2}\int d^{4}x\,e\,(\bar{\Psi}\gamma^{I}e^{\mu}_{I}D_{\mu}\Psi -
    \overline{D_{\mu}\Psi}\gamma^{I}e^{\mu}_{I}\Psi)
\end{equation}

where:

\begin{eqnarray}\label{covariant_derivative}
    D_{\mu}\Psi = \partial_{\mu}\Psi - \frac{1}{4}A_{\mu}^{IJ}\gamma_{I}\gamma_{J}\Psi \\
    \overline{D_{\mu}\Psi} = \partial_{\mu}\bar{\Psi} + \frac{1}{4}\bar{\Psi}\gamma_{I}\gamma_{J}A_{\mu}^{IJ}
\end{eqnarray}

The equation of motion obtained by varying (\ref{action}) with respect to the four dimensional spin
connection $A_{\mu}^{IJ}$ yields:

\begin{equation}\label{connection}
    A_{\mu}^{IJ}=\omega_{\mu}^{IJ} + C_{\mu}^{IJ}
\end{equation}

where $\omega_{\mu}^{IJ}$ is the spin connection compatible with the tetrad $e_{I}^{\mu}$ and
$C_{\mu}^{IJ}$ is the tetrad projection of the contortion tensor:

\begin{equation}\label{contortion}
    C_{\mu}^{IJ} = C_{\mu}^{\nu\delta}e^{I}_{[\nu}e^{J}_{\delta]}
\end{equation}

On solving for $C_{\mu}^{IJ}$ in terms of the fermionic field and inserting the resulting
expression for $A_{\mu}^{IJ}$ in (\ref{action}) one obtains the following:

\begin{equation}\label{action2}
    S[e,\Psi] = \frac{1}{16\pi G}\int
    d^{4}x\,e\,e^{\mu}_{I}e^{\nu}_{J}F^{IJ}_{\mu\nu}[\omega(e)] + \frac{i}{2}\int d^{4}x\,e\,(\bar{\Psi}\gamma^{I}e^{\mu}_{I}D_{\mu}[\omega(e)]\Psi -
    \overline{D_{\mu}[\omega(e)]\Psi}\gamma^{I}e^{\mu}_{I}\Psi) +
    S_{int}[e,\Psi] + S_b
\end{equation}

where $S_{int}$ is the four fermion interaction\footnote{A detailed derivation is included in the
Appendix}:

\begin{equation}\label{int_action}
    S_{int} = -\frac{3}{2}\pi G \frac{\gamma^{2}}{\gamma^{2}+1}
    \int d^{4}x\,e (\bar{\Psi}\gamma_{5}\gamma_{I}\Psi)(\bar{\Psi}\gamma_{5}\gamma^{I}\Psi)
    = -\frac{3}{2}\pi G \frac{\gamma^{2}}{\gamma^{2}+1} \int d^{4}x\,e (j_a^I)^2
\end{equation}

and $S_b$ is a boundary term, given by:

\begin{equation}\label{boundary_term}
    S_b = -\frac{3}{4\kappa\gamma}\oint\limits_{\partial\Sigma}^{} d^3x n_\mu j_a^\mu
\end{equation}

Before we proceed to the $(3+1)$ decomposition of the above action, we write the Dirac action in
terms of Weyl spinors. This will make the decomposition simpler and will also illustrate an
important property of the left and right handed spinors \footnote{In the following we essentially
follow the Appendix of \cite{Thiemann_QSD_V}, filling in some of the steps. We have included this
derivation to make the paper self-contained.}

We expand the second term in (\ref{covariant_derivative})

\begin{eqnarray}\label{connection_decomp}
    A_\mu^{IJ}\gamma_I\gamma_J & = & A_\mu^{i0}\gamma_i\gamma_0 +
    A_\mu^{0i}\gamma_0\gamma_i + A_\mu^{ij} \gamma_i\gamma_j \nonumber \\
    & = & 2A_\mu^{0i}\gamma_0\gamma_i+ A_\mu^{ij}\gamma_i\gamma_j \nonumber  \\
    & = & 2A_\mu^{0i}\left(\begin{array}{cc}-\sigma_i&0\\0&\sigma_i\end{array}\right)
    + iA_\mu^{jk}\epsilon^{ijk}\left(\begin{array}{cc}\sigma_i&0\\0&\sigma_i\end{array}\right)\nonumber \\
    & = & 2i\left(\begin{array}{cc}A_\mu^{i+}\sigma_i&0\\0&A_\mu^{i-}\sigma_i\end{array}\right)
\end{eqnarray}

In the second line we have used the fact that $A_\mu^{IJ}$ is antisymmetric in the internal indices
and that the gamma matrices anticommute. In the third we have used the expressions for the gamma
matrices given in the appendix to expand out the matrix products. In the fourth we have used the
definition of the self and anti-self dual parts of the connection:

\begin{eqnarray}
    A_\mu^{i+} = \frac{1}{2}\epsilon^{ijk}A_\mu^{jk} + iA_\mu^{0i} \nonumber \\
    A_\mu^{i-} = \frac{1}{2}\epsilon^{ijk}A_\mu^{jk} - iA_\mu^{0i}
\end{eqnarray}

Now writing the Dirac spinor $\Psi$ in term of the Weyl spinors $\psi, \eta$, we see that
(\ref{covariant_derivative}) becomes:

\begin{equation}
    D_\mu\Psi = \left(\begin{array}{c}{\cal D}_\mu^+\psi\\{\cal D}_\mu^-\eta\end{array}\right)
\end{equation}

where ${\cal D}_\mu^+\psi=\partial_\mu\psi-\frac{i}{2}A_\mu^{i+}\sigma_i\psi$ and ${\cal
D}_\mu^-\eta=\partial_\mu\eta-\frac{i}{2}A_\mu^{i-}\sigma_i\eta$. Thus the left(right) handed
spinors couple to the self(anti-self) dual parts of the connection.

We now proceed with the (3+1) decomposition of (\ref{action2}).

\section{3+1 decomposition and Legendre Transform}

Consider a spacelike slice $\Sigma$ of the spacetime manifold ${\cal M}$ with unit normal
$n^{\mu}$. Then the Dirac action is:

\begin{eqnarray}\label{dirac_decomp}
    2S_{D} & = & i\int d^{3}x\,dt\,N\,\sqrt{q}\,
    (\bar{\Psi}\gamma_{\mu}{\cal D}_{\nu}\Psi - c.c.)(q^{\mu\nu}-n^{\mu}n^{\nu})\nonumber \\
    & = & i\int d^{3}x\,dt\,N\,\sqrt{q}
    (\bar{\Psi}\gamma^{a}{\cal D}_{a}\Psi + \bar{\Psi}\gamma^0n^{\nu}{\cal D}_{\nu}\Psi - c.c.)\nonumber \\
    & = & i\int d^{3}x\,dt\,N\,\sqrt{q}
    (\psi^\dag\sigma^{a}{\cal D}_{a}^{+}\psi - \eta^\dag\sigma^{a}{\cal D}_{a}^{-}\eta - c.c)
    +\sqrt{q}(t^{\mu}-N^{\mu})(\psi^\dag {\cal D}_{\mu}^{+}\psi + \eta^\dag {\cal D}_{\mu}^{-}\eta -
    c.c) \nonumber \\
    & = & i\int d^{3}x\,dt\,N\,\sqrt{q}
    (\psi^\dag\sigma^{a}{\cal D}_{a}^{+}\psi - \eta^\dag\sigma^{a}{\cal D}_{a}^{-}\eta - c.c) \nonumber \\
    &&+ \sqrt{q}(\psi^\dag\dot{\psi} + \eta^\dag\dot{\eta} -
    \frac{i}{2}A_t^{i\mathbb{C}}\psi^\dag\sigma_i\psi -
    \frac{i}{2}\bar{A}_t^{i\mathbb{C}}\eta^\dag\sigma_i\eta - c.c.) \nonumber \\
    &&- \sqrt{q}N^a(\psi^\dag {\cal D}_a^+\psi + \eta^\dag {\cal D}_a^-\eta - c.c.)
\end{eqnarray}

In the first line we have used the decomposition of the metric $g_{\mu\nu}$ on ${\cal M}$ in terms
on the metric $q_{ab}$ on $\Sigma$ and the unit normal $n^{\mu}$ to $\Sigma$. $q^{\mu\nu}$ projects
tensors and derivatives on ${\cal M}$ to tensors and derivatives on $\Sigma$. ${\cal D}_{\mu}$ and
${\cal D}_{a}$ denote the covariant derivative on ${\cal M}$ and its restriction to $\Sigma$
respectively. In the second line we have used the freedom to fix the gauge in the internal space
such that the contraction of $\gamma_{\mu}$ and $n^{\mu}$ gives us $-\gamma^{0}$. In the third the
decomposition of $n^{\mu}$ in terms of the timelike vector field $t^{\mu}$, the lapse $N$ and the
shift $N^{\mu}$, and the expression of the covariant derivative in terms of the self and anti-self
dual parts of the connection is used. In the last line we have noted that the restriction of
$A_\mu^{i+}$ to $\Sigma$ is the Ashtekar connection $\Gamma_a^i + iK_a^i$. The time component of
$A_\mu^{i+}$ is written as $A_t^{i\mathbb{C}}$ in the fifth line.

Defining $A_t^j := \mathbb{Re}(A_t^{j\mathbb{C}})$ and evaluating the complex conjugate terms
explicitly we get:

\begin{eqnarray}\label{dirac_decomp2}
    S_D & = &  \frac{i}{2}\int d^{3}x\,dt \sqrt{q}(\psi^\dag\dot{\psi} +
    \eta^\dag\dot{\eta} - c.c.) - i\sqrt{q} A_t^{i}(\psi^\dag\sigma_i\psi + \eta^\dag\sigma_i\eta) \nonumber \\
    && - \sqrt{q}N^a(\psi^\dag D_a\psi + \eta^\dag D_a\eta - c.c.) \nonumber \\
    && + N\left[E^a_i(\psi^\dag\sigma^i D_a\psi - \eta^\dag\sigma^i D_a\eta - c.c.)
    + i[K_a,E^a]^k(\psi^\dag\sigma_k\psi + \eta^\dag\sigma_k\eta) \right]
\end{eqnarray}

Here $D_a\psi = \partial_a\psi - \frac{i}{2}\Gamma_a^i\sigma_i\psi$. We can easily see that
contributions of the Dirac action to the gauss, scalar and diffeomorphism constraints are the
coefficients of $A^i_t$, $N$ and $N^a$ respectively. The decomposition of $S_{int}$ is easily done
and we obtain the following form:

\begin{eqnarray}\label{interaction_decomp}
    S_{int} & = &  -\frac{3}{2}\pi G \frac{\gamma^{2}}{\gamma^{2}+1}
                    \int d^{3}x\,dt \sqrt{q}N\left[(\psi^\dag\sigma^a\psi + \eta^\dag\sigma^a\eta)^2 - (-\psi^\dag\psi +
                    \eta^\dag\eta)^2\right]
\end{eqnarray}

From (\ref{dirac_decomp2}) we see that Lagrange multiplier of the matter contribution to the
gravitational gauss constraint is $A_t^j$. In order to get this Lagrange multiplier one must first
start with the 3+1 decomposition of the self-dual gravitational action and then take its real part.
The self-dual gravitational action is:

\begin{equation}\label{self-dual_action}
    S_{SD} = \frac{1}{\kappa}\int d^4x\,e^a_I e^b_J {}^{+}F_{ab}^{IJ}
\end{equation}

${}^{+}F_{ab}^IJ$ is the curvature of the self-dual connection and $e^a_I$ is the usual tetrad.
Doing the 3+1 decomposition in the usual manner yields:

\begin{equation}\label{self-dual_decomp}
    S_{SD} = \frac{1}{\kappa}\int d^3x\,dt\, \left[-i \tilde{E}^b_i\dot{A}_b^i -
    i A_t^{i\mathbb{C}} {\cal D}_b(\tilde{E}^b_i) - i N^a
    tr[F_{ab}\tilde{E}^b] +
    \frac{N}{2\sqrt{q}}tr(F_{ab}[\tilde{E}^a,\tilde{E}^b])\right]
\end{equation}

where $\tilde{E}^b_i$ is the densitized triad, $F_{ab}^i$ is the curvature of the restriction
$A_b^i$ to $\Sigma$ of the complex self-dual connection, and the trace and commutators are taken in
the Lie-algebra of su(2).

Taking the real part of the above action and using the fact that $A_a^i = \Gamma_a^i + iK_a^i$ we
get:

\begin{equation}\label{real_gravitational_action}
    S_{real} = \frac{1}{\kappa}\int
    d^3x\,dt\,\left\{\tilde{E}^b_i \dot{K}^i_b + A^i_t[K_b,\tilde{E}^b]^i
    + 2 N^a D_{[a}K_{b]}^i\tilde{E}^b_i \\
    + \frac{N}{2\sqrt{q}}(R_{ab}^i -[K_a,K_b]^i)[\tilde{E}^a,\tilde{E}^b]_i
    \right\}
\end{equation}

From (\ref{dirac_decomp2}) we see that the momenta conjugate to $\psi$ and $\psi^\dag$ are
$\frac{i}{2}\psi^\dag$ and $-\frac{i}{2}\psi$ respectively.. Then doing the Legendre transform on
$S_{real} + S_D + S_{int}$ we get the following Hamiltonian:

\begin{eqnarray}\label{total_hamiltonian}
    \lefteqn{H_{G+D+int}  =  \int d^{3}x\, A_t^{i}\left\{\frac{1}{\kappa}[K_b,\tilde{E}^b]^i + j^i \right\}}\nonumber \\
     & &{}+N\bigg\{\frac{1}{2\kappa\sqrt{q}}(R_{ab}^i-[K_a,K_b]^i)[\tilde{E}^a,\tilde{E}^b]_i
     +\frac{i}{2\sqrt{q}}\tilde{E}^a_i(\xi^\dag\sigma^i D_a\xi - \rho^\dag\sigma^i D_a\rho - c.c.)\nonumber \\
     & &{}+\frac{1}{2}[K_a,\tilde{E}^a]^k j_k
     - \frac{3}{2} \pi G \frac{\gamma^{2}}{\gamma^{2}+1}[j^2 - (-\xi^\dag\xi + \rho^\dag\rho)^2] + \sqrt{q}\Lambda_0\bigg\} \nonumber \\
     & &{}+N^a\left\{\frac{2}{\kappa}D_{[a}K_{b]}^i\tilde{E}^b_i +
    \frac{i}{2}(\xi^\dag D_a\xi + \rho^\dag D_a\rho - c.c.)\right\}
\end{eqnarray}

where $\xi=q^\frac{1}{4}\psi; \rho=q^\frac{1}{4}\eta$ and $j^i = (\eta^\dag\sigma^i\eta +
\rho^\dag\sigma^i\rho)/2$ is the axial current . We must change variables to make the matter fields
half-densities, because otherwise the connection would become complex \cite{Thiemann_QSD_V}. The
hamiltonian is manifestly a sum of constraints and the form of each constraint is easy to read off
from (\ref{total_hamiltonian}). It is important to note the gravitational Gauss constraint now has
a matter contribution. In the third line we have also added a term coming from the bare
cosmological constant.

\section{Symmetry Reduction and quantization}

We make the ansatz that the background metric is FRW with scale factor $a$. The basic gravitational
variables are:

\begin{eqnarray}\label{FRWMetricVariables}
    E_i^a = a^2\delta^a_i & K_a^i = a^2\dot{a}\delta^a_i & R_{ab}^i = 0
\end{eqnarray}

We assume, for the moment, that the axial current is zero and hence the Gauss constraint is
satisfied. We also assume that the matter contribution to the diffeomorphism constraint is zero.
Later we shall find that these statements are true when we quantize the fermionic field. We are
left with the hamiltonian constraint and this reduces to:

\begin{equation}\label{ReducedHamiltonian}
    {\cal H} = {\cal H}_G + {\cal H}_D + {\cal H}_{int}= -\frac{3}{\kappa}a^3H^2 + a^3\Lambda_0 +
    \frac{i}{a}\left(\xi^\dag\sigma^a\partial_a\xi - \rho^\dag\sigma^a\partial_a\rho\right)
    + \frac{3\kappa}{32a^3}\frac{\gamma^2}{\gamma^2+1}\left[\xi^\dag\xi - \rho^\dag\rho\right]^2 = 0
\end{equation}

where $H = (\dot{a}/a)$ is the Hubble parameter.

We switch to comoving co-ordinates in order to take care of the factor of $1/a$ in ${\cal H}_D$.
${\cal H}_D$ reduces to $i(\xi^\dag\sigma^a\partial_a\xi - \rho^\dag\sigma^a\partial_a\rho)$. We
can then expand $\xi$ and $\rho$ in terms of fourier modes:

\begin{subequations}\label{FermionFourierModes}
\begin{equation}
    \xi(x) = \int \frac{d^3k}{(2\pi)^3} \left\{ \xi_{k\uparrow}e^{-ikx} + \xi_{k\downarrow}e^{-ikx} \right\}
\end{equation}
\begin{equation}
    \rho(x) = \int \frac{d^3k}{(2\pi)^3} \left\{ \xi_{k\downarrow}e^{-ikx} + \xi_{k\uparrow}e^{-ikx} \right\}
\end{equation}
\end{subequations}

where $\xi_{k\uparrow}$ ($\xi_{k\downarrow}$) is a spinor\footnote{The expressions for these
spinors are given in the Appendix} of density weight $1/2$ \footnote{Because as mentioned earlier
the fermionic fields must be half-densities} along the direction $\hat{k}$ in momentum space and
with helicity $1/2$ and $-1/2$ respectively. Thus we can write the quantized field in the usual
manner in term of anticommuting annihilation and creation operators:

\begin{subequations}\label{QuantizedFermionField}
\begin{equation}
    \hat\xi(x) = \int \frac{d^3k}{(2\pi)^3} \left\{ a_k \xi_{k\uparrow} + b^\dag_{-k} \xi_{k\downarrow} \right\} e^{-ikx}
\end{equation}
\begin{equation}
    \hat\rho(x) = \int \frac{d^3k}{(2\pi)^3} \left\{ \bar b_k \xi_{k\downarrow} + \bar a^\dag_{-k} \xi_{k\uparrow} \right\}e^{-ikx}
\end{equation}
\end{subequations}

$\rho$ and $\xi$ are independent fields, therefore we have used $\bar{}$ to distinguish their
operators. These fields satisfy the anticommutation relations:

\begin{subequations}\label{Anticommutation}
\begin{equation}
    \{\hat\xi^\dag_\alpha(x),\hat\xi_\beta(y)\} = \{\hat\rho^\dag_\alpha(x),\hat\rho_\beta(y)\} =
    (2\pi)^3\delta_{\alpha\beta}\delta^3(x,y)\sqrt{q}
\end{equation}
\begin{equation}
    \{\hat\xi_\alpha(x),\hat\xi_\beta(y)\} = \{\hat\xi^\dag_\alpha(x),\hat\rho_\beta(y)\} = 0
\end{equation}
\end{subequations}

The above expressions for the quantized field can be used to easily verify that the spatial current
and the matter contribution to the diffeomorphism constraint are zero, as stated previously. Using
the orthogonality of spinors of opposite helicity, the quantized form of the free Dirac hamiltonian
is easily found to be:

\begin{equation}\label{QuantizedDiracHam}
    \hat{\cal H}_D = \hat{\cal H}_\xi + \hat{\cal H}_\rho = \int \frac{d^3k}{(2\pi)^3}\,|k|
            (a^\dag_k a_k + b^\dag_{-k} b_{-k} + \bar a^\dag_{-k} \bar a_{-k} + \bar b^\dag_k \bar b_k)
\end{equation}

\section{Boguliubov transformation}

The four-fermi interaction is identical to the one which describes the formation of a condensate in
BCS theory\cite{Fetter_Walecka}. Due to this interaction the true vacuum is not the one
corresponding to the Dirac equation but one in which particles and antiparticles of opposite
momenta and helicity are paired\footnote{In BCS theory the pairing happens between particles of
opposite momenta. However, here we have left and right handed fermions therefore the pairing must
include the helicity}. The interacting part is non-diagonal in the present variables. In order to
diagonalize the full matter hamiltonian we have to perform a Boguliubov transformation, which is a
linear canonical transformation to new annihilation and creation operators. We get a new ground
state corresponding to these operators. This BCS ground state is a condensate of Cooper pairs.
Excitations of this "vacuum" are produced by the action of the new operators whose physical effect
is to break up Cooper pairs and produce free fermions and antifermions.

\begin{subequations}\label{BogTrans}
\begin{equation}
    \alpha_k = u_k a_k - v_k b^\dag_{-k}
\end{equation}
\begin{equation}
    \beta_{-k} = u_k b_{-k} + v_k a^\dag_k
\end{equation}
\end{subequations}

Then the new variables $\alpha_k$ and $\beta_{-k}$ satisfy anticommutation relations if $u_k^2 +
v_k^2 = 1$. In terms of the new variables, the old ones are:

\begin{subequations}\label{InverseBogTrans}
\begin{equation}
    a_k = u_k \alpha_k + v_k \beta^\dag_{-k} \\
\end{equation}
\begin{equation}
    b_{-k} = u_k \beta_{-k} - v_k \alpha^\dag_k
\end{equation}
\end{subequations}

In the new variables $\hat{\cal H}_\xi$ becomes:

\begin{equation}\label{FreeTerm}
    \hat{\cal H}_\xi = \int \frac{d^3k}{(2\pi)^3}\, |k| \left(2v_k^2  + (u_k^2 - v_k^2)(m_k + n_{-k}) + 2u_k v_k\Sigma_k\right)
\end{equation}

where $m_k = \alpha^\dag_k\alpha_k$ , $n_{-k} = \beta^\dag_{-k}\beta_{-k}$ are the new number
operators and $\Sigma_k = \alpha^\dag_k\beta^\dag_{-k} + \beta_{-k}\alpha_k$ is the off-diagonal
part.

\section{Four-fermion term}

The interaction hamiltonian is an attractive four-fermion term which causes the formation of the
fermion condensate. In the this section we use the mode expansion for the fermion field to expand
this term and then apply the Boguliubov transformation to it.

The four-fermion term is:

\begin{eqnarray}\label{FourFermionTerm}
    \hat H_{int} &=& \frac{3\kappa}{32a^3}\frac{\gamma^2}{\gamma^2+1}
            \int d^3x\,\left(\hat\xi^\dag\hat\xi - \hat\rho^\dag\hat\rho\right)^2 \nonumber \\
    &=&  \frac{\alpha}{a^3} \int d^3x\, \left(\hat\xi^\dag\hat\xi\hat\xi^\dag\hat\xi +
                \hat\rho^\dag\hat\rho\hat\rho^\dag\hat\rho
                - \hat\rho^\dag\hat\rho\hat\xi^\dag\hat\xi - \hat\xi^\dag\hat\xi\hat\rho^\dag\hat\rho
                \right) \nonumber \\
    &=& \hat H_1 + \hat H_2 + \hat H_{\rho\xi} + \hat H_{\xi\rho}
\end{eqnarray}

where $\alpha = \frac{3\kappa}{32}\frac{\gamma^2}{\gamma^2+1}$. Now we can write $\hat\rho$ as:

\begin{equation}\label{rho}
    \hat\rho(x) = \int \frac{d^3k}{(2\pi)^3} \left\{\bar b_{-k} \xi_{k\uparrow}+\bar a^\dag_{k} \xi_{k\downarrow}\right\}e^{ikx}
\end{equation}

by doing changing variables from $k$ to $-k$ in the integration. Then by comparing (\ref{rho}) and
(\ref{QuantizedFermionField}a) we see that one can switch from $\hat\xi$ to $\hat\rho$ (or vice
versa) by changing $a_k \leftrightarrow \bar b_{-k}$.

Now using the anticommutation relations for the fermionic fields we can write $\hat H_1$ as:

\begin{equation}
    \hat H_1 = \alpha \int d^3x\, \hat\xi^\dag \hat\xi +
    \frac{\alpha}{a^3} \int d^3x\,
    \hat\xi^\dag_\alpha\hat\xi^\dag_\beta\hat\xi^\beta\hat\xi^\alpha
    = \hat N_\xi + \hat H_{\xi\xi}
\end{equation}

Using (\ref{QuantizedFermionField}a) and (\ref{BogTrans}) $\hat N_\xi$ becomes:

\begin{equation}\label{xiNumberOp}
    \hat N_\xi = \alpha \int \frac{d^3k}{(2\pi)^3}\, \left[a^\dag_k a_k - b^\dag_{-k} b_{-k} \right]
                = \alpha \int \frac{d^3k}{(2\pi)^3}\, \left[m_k - n_{-k} \right]
\end{equation}

Likewise for $\hat\rho$ we have:

\begin{equation}
    \hat H_2 = \alpha \int d^3x\, \hat\rho^\dag\rho +
    \frac{\alpha}{a^3} \int d^3x\,
    \hat\rho^\dag_\alpha\hat\rho^\dag_\beta\hat\rho^\beta\hat\rho^\alpha
    = \hat N_\rho + \hat H_{\rho\rho}
\end{equation}

and $\hat N_\rho$ is:

\begin{equation}\label{rhoNumberOp}
    \hat N_\rho = \alpha \int \frac{d^3k}{(2\pi)^3}\, \left[\bar b^\dag_k \bar b_k - \bar a^\dag_{-k} \bar a_{-k} \right]
                = \alpha \int \frac{d^3k}{(2\pi)^3}\, \left[\bar n_k - \bar m_{-k} \right]
\end{equation}

To explicitly evaluate $\hat H_{\xi\xi}$ and $\hat H_{\rho\xi}$ we use the mode expansion
(\ref{QuantizedFermionField}) and the anticommutation relations of the fermionic operators. Then
$\hat H_{\rho\rho}$ and $\hat H_{\xi\rho}$ are obtained by simply using the substitution $ a_k
\leftrightarrow b_{-k}$. After some algebra we obtain the following expression:

\begin{multline}\label{InteractionTerm}
    \hat H_{\xi} + \hat H_{\xi\xi} + \hat N_{\xi} =
    \int \frac{d^3k}{(2\pi)^3} |k|(a^\dag_k a_k + b^\dag_{-k} b_{-k}) -
    \alpha  \int \frac{d^3k d^3p}{(2\pi)^6}\,
    2 a^\dag_k a_k \left[
    \left(\xi^\dag_{k\uparrow}\xi_{p\downarrow}\right)\left(\xi^\dag_{p\downarrow}\xi_{k\uparrow}\right)
    +{}1 \right] \\
    -{}\alpha  \int \frac{d^3k d^3p d^3k'd^3p'}{(2\pi)^6}\, \delta^3(k+k'-p-p')
    \bigg\{
    a^\dag_k a^\dag_{k'} a_p a_{p'}\left(\xi^\dag_{k\uparrow}\xi_{p\uparrow}\right)\left(\xi^\dag_{k'\uparrow}\xi_{p'\uparrow}\right)
    + \\
    b^\dag_{-k} b^\dag_{-k'} b_{-p} b_{-p'} \left(\xi^\dag_{p\downarrow}\xi_{k\downarrow}\right)\left(\xi^\dag_{p'\downarrow}\xi_{k'\downarrow}\right)
    \bigg\} + \delta^3(k-k'+p-p')
    a^\dag_k b^\dag_{-k'} b_{-p} a_{p'} \bigg\{
    \left(\xi^\dag_{k\uparrow}\xi_{k'\downarrow}\right)\left(\xi^\dag_{p\downarrow}\xi_{p'\uparrow}\right)
    \\
    + \left(\xi^\dag_{k\uparrow}\xi_{p'\uparrow}\right)\left(\xi^\dag_{p\downarrow}\xi_{k'\downarrow}\right)
    \bigg\} + \alpha  \int \frac{d^3k}{(2\pi)^3}\, \left[a^\dag_k a_k - b^\dag_{-k} b_{-k} \right]\\
    = \hat H_{\xi} + \alpha  \int \frac{d^3k d^3p}{(2\pi)^6}\, A_0 V_0
    -\alpha  \int \frac{d^3k d^3p d^3k'd^3p'}{(2\pi)^6}\, \big[\delta^3(k+k'-p-p')(A_1 V_1 + A_2 V_2) +  \\
    \delta^3(k-k'+p-p') A_3 V_3 \big] + \hat N_{\xi}
\end{multline}

In the last line the $A${\small s} denote the operator products and the $V${\small s} denote the spinor products. Also
in the above expression and henceforth we only use dedensitized spinors. There is a factor of $a^3$ in front of the
whole expression which we set to $1$ for now. The factor is re-introduced later when appropriate.

Now using momentum conservation we can simplify $A_1$ as follows.

\begin{equation}
    A_{1} = a^\dag_k a^\dag_{k'} a_p a_{p'} = a^\dag_k a^\dag_{k'} a_{k-q} a_{k'+q}
\end{equation}

Using Wick's theorem and the operator identities in the Appendix the above expression can be
written as:

\begin{eqnarray}\label{NormalOrderedOps}
    A_{1} &=& N(A_{1})+ \bigg\{ - N(a^\dag_k a_{k-q})\wick{1}{<1a^\dag_{k'}>1a_{k'+q}}
    - N(a^\dag_{k'} a_{k'+q})\wick{1}{<1a^\dag_{k}>1a_{k-q}}
    + N(a_k^\dag a_{k'+q}) \wick{1}{<1a^\dag_{k'}>1a_{k-q}} \nonumber \\
    &&{} +N(a_{k'}^\dag a_{k-q})\wick{1}{<1a^\dag_{k}>1a_{k'+q}}
    - \wick{1}{<1a^\dag_{k}>1a_{k-q}} \wick{1}{<1a^\dag_{k'}>1a_{k'+q}}
    + \wick{1}{<1a^\dag_{k}>1a_{k'+q}} \wick{1}{<1a^\dag_{k'}>1a_{k-q}} \bigg\}\nonumber \\
    &=&N(A_{1}) +\bigg\{- N(a^\dag_k a_k) v^2_{k'} \delta_{q,0} - N(a^\dag_{k'} a_{k'}) v^2_k \delta_{q,0}
            + N(a^\dag_k a_k) v^2_{k'}\delta_{k',k-q} + \nonumber \\
         &&   N(a^\dag_{k'} a_{k'}) v^2_k\delta_{k',k-q} - v^2_k v^2_{k'} \delta_{q,0} + v^2_k v^2_{k'} \delta_{k',k-q}
            \bigg\}
\end{eqnarray}

Inserting the above expression for $A_1$ into (\ref{InteractionTerm}) and integrating first over
the delta function in (\ref{InteractionTerm}) and then over the delta functions in
(\ref{NormalOrderedOps}) we obtain after relabelling some indices and some algebraic manipulations
we have:

\begin{multline}\label{Int1}
  - \alpha  \int \frac{d^3k d^3k' d^3q}{(2\pi)^9}\, A_1 V_1(k,k',q)
            = \\-N(V_1) + \alpha  \int \frac{d^3k d^3k'}{(2\pi)^6}\,
    \left[ N(a^\dag_k a_k)v^2_{k'} + N(a^\dag_{k'}a_{k'})v^2_k + v^2_k v_{k'}^2 \right]
    \left[ 1 -
    \left(\xi^\dag_{k\uparrow}\xi_{k'\uparrow}\right)\left(\xi^\dag_{k'\uparrow}\xi_{k\uparrow}\right)
    \right]
\end{multline}

where $N(V_1)$ is quartic in the creation and annihilation operators.

$A_2$ can be dealt with in a similar manner and after some computations we find:

\begin{multline}\label{Int2}
    - \alpha  \int \frac{d^3k d^3k' d^3q}{(2\pi)^9}\, (A_1 V_1 + A_2 V_2) = \\
    -N(V_1 + V_2) - \alpha  \int \frac{d^3k d^3k'}{(2\pi)^6}\, 2
    \left[N(a^\dag_k a_k) + N(b^\dag_{-k}b_{-k}) + v^2_k \right] v^2_{k'}
    \left[
    \left(\xi^\dag_{k\uparrow}\xi_{k'\uparrow}\right)\left(\xi^\dag_{k'\uparrow}\xi_{k\uparrow}\right)
    -1
    \right]
\end{multline}

The term with $A_3$ yields:

\begin{multline}\label{Int3}
    - \alpha  \int \frac{d^3k d^3k' d^3q}{(2\pi)^9}\, A_3 V_3 =
    - N(V_3) - \bigg \{
    \left[N(a^\dag_k a_k) + N(b^\dag_{-k}b_{-k}) + v^2_k \right]v^2_{k'}
    \left[
    \left(\xi^\dag_{k\uparrow}\xi_{k'\downarrow}\right)\left(\xi^\dag_{k'\downarrow}\xi_{k\uparrow}\right) + 1
    \right] \\
    + \left[N(a^\dag_k b^\dag_{-k}) + N(b_{-k}a_k) + u_k v_k \right]u_{k'}v_{k'}
    \Re \left[ \left(\xi^\dag_{k\uparrow}\xi_{k'\uparrow}\right)\left(\xi^\dag_{k'\downarrow}\xi_{k\downarrow}\right)
    \right] \bigg\}
\end{multline}

Above we have dealt with the terms of $\hat H_{\xi\xi}$. Doing similar manipulations with $\hat
H_{\xi\rho}$ we find:

\begin{equation}\label{IntCross}
    \hat H_{\xi\rho} = \hat H_{\rho\xi} = - \alpha  \int \frac{d^3k d^3k'}{(2\pi)^6}\,
    \left( a^\dag_k a_k - b^\dag_{-k} b_{-k} + \bar b^\dag_{-k} \bar b_{-k} - \bar a^\dag_k \bar a_k \right)
\end{equation}

In the above equation we have a seemingly divergent integral over the momenta $k'$. This is dealt
with by imposing a momentum cutoff. We get:

\begin{equation*}
    \int \frac{d^3k}{2\pi^3} = \frac{1}{2\pi^2}\int k^2 dk
    = \frac{1}{2\pi^2}\int\limits_{0}^{\hbar \omega_D} E^2 dE =
    \frac{(\hbar\omega_D)^3}{6\pi^2}=C_1
\end{equation*}

The sum of (\ref{FreeTerm}), (\ref{Int1}), (\ref{Int2}), (\ref{Int3}) and the first half of (\ref{IntCross}) gives us
the matter hamiltonian corresponding only to the field $\xi$. The other half corresponding to $\rho$ can be is
identical except for the substitution $a_k \leftrightarrow \bar b_{-k}$.

\begin{multline}
    \hat H(\xi) =  \int \frac{d^3k}{(2\pi)^3} \left( a^\dag_k a_k + b^\dag_{-k} b_{-k}\right) \left( |k| - C_1\alpha \right)
        - \alpha  \int \frac{d^3k}{(2\pi)^3}\bigg\{ \left[ N(a^\dag_k b^\dag_{-k}) + N(b_{-k} a_k) + u_k v_k \right] u_{k'} v_{k'} V'_1  \\
        + \left[ N(a^\dag_k a_k) + N(b^\dag_{-k} b_{-k}) + v^2_k \right] v^2_{k'} V'_2 \bigg\} \\
        =  \int \frac{d^3k}{(2\pi)^3} \left[ (u^2_k - v^2_k)(m_k + n_{-k}) + 2 u_k v_k \Sigma_k + 2v^2_k \right](|k|- C_1\alpha)
         - \alpha  \int \frac{d^3k d^3k'}{(2\pi)^6} \bigg\{ \big[(u^2_k - v^2_k)\Sigma_k \\
         + 2 u_k v_k (m_k + n_{-k}) + u_k v_k \big] u_{k'} v_{k'}V'_1
        + \left[ (u^2_k - v^2_k)(m_k + n_{-k}) + 2 u_k v_k \Sigma_k + v^2_k \right] v^2_{k'} V'_2
        \bigg\}
\end{multline}

where:

\begin{equation}
    \Sigma_k = \alpha^\dag_k \beta^\dag_{-k} - \beta_{-k} \alpha_k  \nonumber
\end{equation}

and,

\begin{subequations}
\begin{equation}
    V'_1(k,k') = \mathbb{Re}\left\{(\xi^\dag_{k\uparrow}\xi_{k'\uparrow})(\xi^\dag_{k'\downarrow}\xi_{k\downarrow}) \right\}
\end{equation}
\begin{equation}
    V'_2(k,k') = (\xi^\dag_{k\uparrow}\xi_{k'\uparrow})(\xi^\dag_{k'\uparrow}\xi_{k\uparrow}) +
            (\xi^\dag_{k\uparrow}\xi_{k'\downarrow})(\xi^\dag_{k'\downarrow}\xi_{k\uparrow}) = 1
\end{equation}
\end{subequations}

Where in the second line we have used the expressions for spinors given in the Appendix. Now we can
easily apply the Boguliubov transformation to the above hamiltonian and then collect terms
according to their operator coefficients. This process yields:

\begin{multline}\label{FullMatterHam}
    \hat H(\xi) = -\hat N(V) +  \int\frac{d^3k}{(2\pi)^3} \bigg\{ (m_k + n_{-k})\left[ (u^2_k - v^2_k)E_k + 2u_k v_k \Delta_k
    \right] + \Sigma_k \left[ 2u_k v_k E_k - (u^2_k - v^2_k)\Delta_k \right] \\
    + \left[ 2v^2_k E_k + E'v^2_k - u_k v_k \Delta_k \right] \bigg\}
     = -\hat N(V) + \hat K_1 + \hat K_2 + \hat U
\end{multline}

where:

\begin{subequations}\label{Defs}
\begin{equation}
    E' = \alpha \int \frac{d^3k'}{(2\pi)^3} \frac{v_{k'}^2}{2} V'_2(k,k')
\end{equation}
\begin{equation}
    E_k =  |k| - C_1\alpha - E'
\end{equation}
\begin{equation}
    \Delta_k = \alpha \int \frac{d^3k'}{(2\pi)^3} V'_1(k,k') u_{k'} v_{k'}
\end{equation}
\end{subequations}

In order to make the full matter hamiltonian diagonal we set the coefficient of $\Sigma_k$ in
(\ref{FullMatterHam}) to zero. This allows us to solve for $u_k$ and $v_k$ in terms of $\Delta_k$
and $E_k$.

Since $u_k^2 + v_k^2=1$, it is natural to use trigonometric variables. We set $u_k = \cos\theta$
and $v_k = \sin\theta$. Then we have:

\begin{eqnarray}\label{GapEqn1}
    &&2 u_k v_k E_k - (u_k^2 - v_k^2)\Delta_k = 0 \\
    \Rightarrow && \sin(2\theta) E_k = \cos(2\theta) \Delta_k \nonumber \\
    \Rightarrow && \tan(2\theta) = \frac{\Delta_k}{E_k}, \quad
    \sin(2\theta) = \frac{\Delta_k}{\sqrt{\Delta^2_k + E^2_k}} = \frac{\Delta_k}{\epsilon_k}, \quad
    \cos(2\theta) = \frac{E_k}{\epsilon_k}
\end{eqnarray}

Using the above the various terms in (\ref{FullMatterHam}) become:

\begin{subequations}\label{FinalMatterHam}
 \begin{equation}\label{PotentialEnergy}
   \hat U =  \int \frac{d^3k}{(2\pi)^3} \left\{ \left(1 - \frac{E_k}{\epsilon_k}\right)\left(E_k + \frac{E'_k}{2}\right) - \frac{\Delta_k^2}{2 \epsilon_k}  \right\}
\end{equation}
\begin{equation}
    \hat K_1 =  \int \frac{d^3k}{(2\pi)^3} \epsilon_k(m_k + n_{-k})
\end{equation}
\end{subequations}

From (\ref{FinalMatterHam}b) it is clear that the spectrum is now bounded from below by $\Delta_k$
which therefore is the mass gap.

Now we do a rough calculation to estimate the value of $\Delta_k$. First let:

\begin{equation}
    u_k = (\frac{1}{2} + x_k)^\frac{1}{2} \quad v_k = (\frac{1}{2} - x_k)^\frac{1}{2}
\end{equation}

Then (\ref{GapEqn1}) becomes:

\begin{eqnarray}
    && 2E_k(\frac{1}{4} - x_k^2)^\frac{1}{2} - 2x_k \Delta_k = 0 \nonumber \\
    \Rightarrow && x_k = \pm\frac{E_k}{2\sqrt{E_k^2 + \Delta_k^2}} \nonumber \\
 \end{eqnarray}

Inserting the solution for $x_k$ into the expression (\ref{Defs}c) for $\Delta_k$, we get the gap
equation:

\begin{equation}\label{GapEqn2}
    \Delta_k = \alpha \int \frac{d^3k'}{(2\pi)^3} \frac{V'_1(k,k')\Delta_{k'}}{2\sqrt{E_{k'}^2 + \Delta_{k'}^2}}
\end{equation}

In the above expression the potential  $ V'_1 \sim \mathcal{O}(1)$. We use a mean-field approximation to set the value
of this potential to a constant $V_a$.

Then (\ref{GapEqn2}) becomes:

\begin{eqnarray}\label{GapEqn3}
    && \Delta \approx \alpha V_a D(0) \int\limits_{-\hbar\omega_D}^{\hbar\omega_D} dE_k \frac{\Delta}{2\sqrt{E_k^2 +
    \Delta^2}} \nonumber \\
    \Rightarrow && 1 \approx \frac{\alpha V_a D(0)}{2}
    \ln\frac{\sqrt{(\hbar\omega_D)^2 + \Delta^2} + \hbar\omega_D}{\sqrt{(\hbar\omega_D)^2 + \Delta^2} - \hbar\omega_D} \nonumber \\
    \Rightarrow && \Delta \approx \frac{2\hbar\omega_D \exp^{\frac{\nu}{2}}}{\exp^{\nu} - 1} \qquad \left(\nu = \frac{2}{\alpha V_a D(0)} \right)
\end{eqnarray}

where in the first line we have used the fact that $\frac{d^3k}{2\pi^3} \approx D(0) dE_k$. $D(0)$
is the density of states at the fermi surface. We note that the gap depends on the Immirzi
parameter which is contained in $\alpha$. Now the density of states, $D(E)$, for a field in a
3-dimensional box of volume $V$ is:

\begin{equation}
     D(E) = \frac{d\,N}{d\,E} = \frac{d\,N}{d\,K}\frac{d\,K}{d\,E} = \frac{VE^2}{\pi^2}
\end{equation}

\begin{figure}[htb]
\includegraphics[scale=0.4]{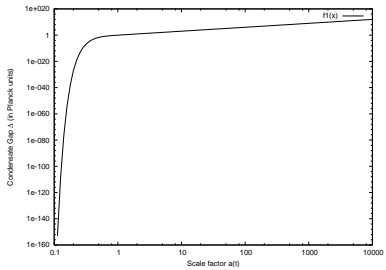}
\caption{\label{fig:gap} Condensate gap as a function of the scale factor}
\end{figure}

The volume of our co-moving box, and hence $D(E)$, scales as $a^3$. Therefore the gap is an
increasing function with respect to $t$. This behavior is shown in Fig \ref{fig:gap}.

The gap has different behavior in the strong ($V_a D(0) >> 1$) and weak ($V_a D(0) << 1$)coupling
limits, corresponding to $a >> 1$ and $a << 1$.

\begin{eqnarray}
    \Delta \sim 2\hbar\omega_D \exp^{\frac{-1}{\alpha V_a D(0)}}  \quad \mbox{Weak coupling} \nonumber \\
    \Delta \sim 2\hbar\omega_D \alpha V_a D(0) \quad \mbox {Strong coupling} \nonumber \\
\end{eqnarray}

For small $a$ the gap is exponentially suppressed and for large $a$ it grows as $a^3$. In
particular, in an inflating background the gap grows as $\exp^{3Ht}$.

One has to keep in mind that this is a semiclassical calculation and breaks down for small $a$ as
we enter a non-perturbative regime where quantum gravitational fluctuations of the metric must be
taken into account.

\section{Discussion}

Eqn. (\ref{PotentialEnergy}) is the expression for the potential energy of the fermi gas. The gap
equation (\ref{GapEqn2}) has two solutions. The trivial solution is zero and corresponds to the
free fermi gas. In this case (\ref{PotentialEnergy}) reduces to the Hartree-Fock potential energy
for the free fermi gas \cite{Fetter_Walecka}. When the condensate forms the potential is reduced by
the amount given by the last term in (\ref{GapEqn2})\footnote{The other terms in (\ref{GapEqn2})
are also affected when he have a condensate. However, this is perturbation is negligible compared
to the that due to the gap term}. The full Hamiltonian constraint (\ref{ReducedHamiltonian}) now
becomes:

\begin{equation}\label{NewHamiltonian}
    \frac{1}{V}\int d^3x \,a^3 \,\mathcal{H} = -\frac{3}{\kappa}a^3H^2 + a^3(\Lambda_0 - \Lambda_{corr})
        + a^3 \int \frac{d^3k}{(2\pi)^3} \sqrt{E_k^2 + \Delta_k^2}(m_k + \bar m_k + n_{-k} + \bar n{-k})
\end{equation}

where $V$ is the volume of integration over the three-manifold. The correction to the cosmological constant is given by
two times the last term of (\ref{PotentialEnergy})\footnote{We have a contribution from the left and right handed
spinors} :

\begin{equation}\label{EffCC}
    \Lambda_{corr} = 2\int \frac{d^3k}{(2\pi)^3} \frac{\Delta_k^2}{2 \sqrt{E_k^2 + \Delta_k^2}}
    \approx \frac{2\Delta^2}{\alpha}
\end{equation}

where the third expression is obtained by using the approximation discussed at the end of the
previous section. In \cite{Alexander2004,Alexander} a perturbative one-loop calculation done for fermions
coupled to gravity via a quartic potential showed that the cosmological constant must be
proportional to $\Delta^2$. Here we have done a non-perturbative calculation to demonstrate that
this expectation is indeed borne out albeit it is the correction $\Lambda_{corr}$, and not
$\Lambda_0$, which is proportional to $\Delta^2$.


\section{Conclusion}

In this work we have demonstrated that when a covariant coupling to fermions in General relativity
induces a four fermion coupling, the Hamiltonian reduces to a BCS theory.  The gravitational field
also induces a chemical potential which creates a Fermi-surface.  By employing the appropriate
Boguliobov transformation we were able to diagonalize this Hamiltonian and evaluate the energy gap.
This gap played the role of negating the cosmological constant. In a time dependent background the
gap is also time dependent. However, further analysis is needed to make this expectation concrete
and we leave this up to a future work. By extending our mechanism to the full theory and
incorporating fermionic spin-networks (open holonomies) we hope to show that the BCS mechanism
persists in the semiclassical approximation.

There are questions this work raises that need to be explored further:
\begin{itemize}
    \item All fermionic species, regardless of whether they couple to Yang-Mills or electromagnetic fields,
also couple to gravity. Therefore a four-fermi interaction is induced for all fermions and does not
distinguish between different species. In the real world we have fermions that condense (quarks)
and those that are free (electrons and neutrinos). It remains to be understood how in this
mechanism can one include interactions which would distinguish between different species, allowing
some to condense and others to remain free and we will pursue this in a forthcoming paper \cite{Alexander3}.
    \item How would the generic inflationary scenario be modified due to the presence of the gap?
$<\bar\Psi\Psi>$ develops a non-zero vev and is a scalar \cite{Prokopec:2006yh}. Can this composite scalar then play the
role of the scalar field in cosmology?
    \item We have not studied the effects of gravitational perturbations on the condensate. In particular if $\Gamma^i_a$ is non-zero then the
number operator (\ref{xiNumberOp}) would be modified by a term proportional to $e^a_i \Gamma_a^i$.
This would increase the chemical potential thereby decreasing $\Lambda_{corr}$. The effect of these
perturbations and the other questions mentioned above will be studied in a future work.
\end{itemize}

\begin{acknowledgments}
We would like to thank Leon Cooper, Robert Brout  and Michael Peskin for illuminating the subtelties of the BCS theory.  Some aspects (in the context of De Sitter space) of this work was initiated in a collaboration with Felipe Llanes-Estrada and Richard Hill.   DV would like to thank Florian Conrady and Arif Mohd for many useful and enlightening discussions.
\end{acknowledgments}

\appendix*
\section{}

\subsection{Weyl Representation}

For the internal space we use the metric with signature $(-+++)$. For this signature, the gamma
matrices in the Weyl representation are:

\begin{equation}\label{gamma_matrices}
    \gamma^{0}=\left(\begin{array}{cc}0&1\\-1&0\end{array}\right);
    \gamma^{a}=\left(\begin{array}{cc}0&\sigma^{a}\\\sigma^{a}&0\end{array}\right);
    \gamma^{5}=\left(\begin{array}{cc}-1&0\\0&1\end{array}\right);
\end{equation}

\subsection{Spinors}

The expressions for the dedensitized spinors are:

\begin{equation}\label{spinors}
    \xi_{k\uparrow} = \left( \begin{array}{c} \cos\frac{\theta}{2} \\ \sin\frac{\theta}{2}\,\exp^{i\phi} \end{array} \right)
    \,\,\,
    \xi_{k\downarrow} = \left( \begin{array}{c} -\sin\frac{\theta}{2} \\ \cos\frac{\theta}{2}\,\exp^{i\phi} \end{array} \right)
\end{equation}

It is manifest the above spinors are orthogonal and by changing $k$ to $-k$ one can check the useful identity
$\xi_{k\uparrow} = - \xi_{-k\downarrow}$. A factor $q^{\frac{1}{4}}$ is required to convert the above spinors into
half-densities.

\subsection{Certain operator contractions}

Following are operator identities used in the main text:

\begin{eqnarray}\label{BogIdentities}
    m_k = \alpha^\dag_k \alpha_k;\,\,\,  n_{-k} = \beta^\dag_{-k}
    \beta_{-k}; \,\,\,
     \Sigma_k = \alpha^\dag_k \beta^\dag_{-k} - \beta_{-k}\alpha_k \\
    N(a^\dag_k a_k) = u^2_k m_k - v_k^2 n_{-k} + u_k v_k \Sigma_k \\
    N(b^\dag_{-k}b_{-k}) = u^2_k n_{-k} - v_k^2 m_k + u_k v_k \Sigma_k \\
    N(b_{-k} a_k) = u_k^2 \beta_{-k}\alpha_k  - v^2_k \alpha^\dag_k\beta^\dag_{-k} - u_k v_k (m_k + n_{-k}) \\
    N(a^\dag_k b^\dag_{-k}) = u_k^2 \alpha^\dag_k\beta^\dag_{-k} - v^2_k \beta_{-k}\alpha_k - u_k v_k (m_k + n_{-k}) \\
    \wick{1}{<1a^\dag_k >1a_{k'}} = \wick{1}{<1b^\dag_{-k}>1b_{-k'}} = v_k^2 \delta_{k,k'} ; \,\,\,
    \wick{1}{<1b_{-k}>1a_{k'}} = \wick{1}{<1a^\dag_k>1b^\dag_{-k'}} = u_k v_{k'} \delta_{k,k'}
\end{eqnarray}

\begin{subequations}\label{Potentials}
\end{subequations}

\subsection{Derivation of the torsion term}

The following is essentially the content of \cite{Rovelli_Perez_1}. We include this derivation here
in order to keep this paper self-contained. The Holst action can be written as:

\begin{equation}
    S_{H} = \frac{1}{2\kappa}\int d^{4}x\,e\,e^{\mu}_{I}e^{\nu}_{J}P^{IJ}{}_{KL}F^{KL}_{\mu\nu}
\end{equation}

where:

\begin{equation}
    P^{IJ}{}_{KL} = \delta^I_{[K}\delta^J_{L]} - \frac{1}{2\gamma}\epsilon^{IJ}{}_{KL}
\end{equation}

whose inverse is:

\begin{equation}
    P^{-1}{}_{IJ}{}^{KL} = \frac{\gamma^2}{\gamma^2+1}\left( \delta^K_{[I}\delta^L_{J]} - \frac{1}{2\gamma}\epsilon_{IJ}{}^{KL}\right)
\end{equation}

Variation of the Holst action w.r.t the connection yields:

\begin{equation}
    \frac{\delta S_H}{\delta A_\nu^{KL}} = -\frac{1}{\kappa}D_\mu \left(
    e\,e^{[\mu}_{I}e^{\nu]}_{J}\right)P^{IJ}{}_{KL}
\end{equation}

Likewise variation of the Dirac action w.r.t yields:

\begin{eqnarray}
    \frac{\delta S_D}{\delta A_\nu^{KL}} &=& - \frac{\i}{8}e\bar\Psi\{\gamma_{[K}\gamma_{L]},\gamma^I\}e^\nu_I\Psi \nonumber\\
            &=& \frac{e}{4}\epsilon^I{}_{KLM}(\bar\Psi\gamma_5\gamma^M\Psi)e^\nu_I
\end{eqnarray}

In the second line we have used the identity: $\{\gamma_{[K}\gamma_{L]},\gamma_I\} =
2i\epsilon_{IKLM}\gamma_5\gamma^M$. Therefore the variation of the action $S_H + S_D$ w.r.t to the
connection yields:

\begin{equation}\label{Gauss_Law}
    D_\mu \left(e\,e^{[\mu}_{I}e^{\nu]}_{J}\right)P^{IJ}{}_{KL} = \frac{\kappa e}{4}\epsilon^I{}_{KLM}j_a^M e^\nu_I
\end{equation}

where $j_a^M$ is the axial current given by $\bar\Psi\gamma_5\gamma^M\Psi$.

Writing the connection as $A_\mu^{IJ} = \omega_\mu^{IJ} + C_\mu^{IJ}$ where $\omega$ is the
connection compatible with the tetrad, and using $P^{-1}{}^{IJ}{}_{KL}$ (\ref{Gauss_Law}) becomes:

\begin{equation}
    C_{\mu[P}{}^\mu e^\nu_{Q]} + C_{[PQ]}{}^\nu =
    \frac{\kappa}{4}\frac{\gamma^2}{\gamma^2+1}e^\nu_I j_{a\,M}\left\{ \epsilon^{MI}{}_{PQ}
            + \frac{1}{\gamma}\delta^M_{[P}\delta^I_{Q]} \right\}
\end{equation}

Tracing over $\nu$ and $P$ we obtain:

\begin{equation}
    C_{\mu Q}{}^\mu = \frac{3}{8}j_{a\,Q}\frac{\gamma}{\gamma^2 + 1}
\end{equation}

From the above two equation we obtain:

\begin{equation}
    C_{[PQ]R} = \frac{\kappa}{4}\frac{\gamma^2}{\gamma^2 + 1}j^M_a \left\{ \epsilon_{MPQR}
            - \frac{1}{2\gamma}\delta_{M[P}\delta_{Q]R}\right\}
\end{equation}

where we raise and lower indices using the tetrad. Then we have:

\begin{equation}
    C_{PQR} = C_{[PQ]R} + C_{[RP]Q} + C_{[QR]P}
\end{equation}

and finally:

\begin{equation}
    C_\mu^{IJ} = \frac{\kappa}{4}\frac{\gamma^2}{\gamma^2 + 1}j^M_a\left \{ \epsilon_{MK}{}^{IJ}e^K_\mu
            - \frac{1}{2\gamma}\delta^{[J}_M e^{I]}_\mu \right\}
\end{equation}

Inserting the above expression into the first-order gravity+matter action yields (\ref{action2}).
The contribution comes only from $S_D$. The Holst action yields the boundary term
(\ref{boundary_term}).

\bibliography{chiral}

\begin{thebibliography}{1}
\bibitem{robert}
  R.~H.~Brandenberger and A.~R.~Zhitnitsky,
  Phys.\ Rev.\ D {\bf 55}, 4640 (1997)
  [arXiv:hep-ph/9604407].

\bibitem{nima}
  N.~Arkani-Hamed, P.~Creminelli, S.~Mukohyama and M.~Zaldarriaga,
  JCAP {\bf 0404}, 001 (2004)
  [arXiv:hep-th/0312100].


\bibitem{Rovelli_Perez_1}
A. Perez and C. Rovelli, Physical Review D {\bf 73},  044013  (2006).

\bibitem{Thiemann_QSD_V}
T. Thiemann, Classical and Quantum Gravity {\bf 15},  1281  (1998).

\bibitem{Fetter_Walecka}
A. Fetter and J.~D. Walecka, {\em Quantum theory of many-particle systems}
  (Dover, Mineola, N.Y., 2003), Chap.~10, p.\ 326.

\bibitem{Prokopec:2006yh}
  T.~Prokopec,
  arXiv:gr-qc/0603088.
\bibitem{Alexander2004}
S. Alexander, M. Mbonye, and J. Moffat,   (2004).
\bibitem{Alexander3}
S. Alexander, D. Vaid, To appear
\bibitem{Alexander}
  S.~Alexander,
  Phys.\ Lett.\ B {\bf 629}, 53 (2005)
  [arXiv:hep-th/0503146].

\end{thebibliography}

\end{document}